\documentclass[
  journal=pasa,
  manuscript=research-paper, 
  year=2023,
  volume=42,
]{cup-journal}

\usepackage{microtype,siunitx,booktabs}
\usepackage{color,soul}
\usepackage{amsmath}
\usepackage{lscape}
\usepackage{multirow}
\usepackage{comment}
\usepackage{IEEEtrantools}
\usepackage[colorlinks=true,linkcolor=blue, citecolor=blue, urlcolor=blue]{hyperref}
\usepackage[nolist]{acronym}
\usepackage{subcaption}
\usepackage{flushend}

\newcommand\norm[1]{\left\lVert#1\right\rVert}

\title{Mitigation of self-generated RFI using ASKAP's phased array feeds}

\author{L. Louren\c{c}o}
\affiliation{Sydney Institute for Astronomy, School of Physics, University of Sydney, Sydney, New South Wales 2006, Australia.}
\alsoaffiliation{CSIRO Space and Astronomy, PO Box 76, Epping, New South Wales 1710, Australia}
\email[L. Louren\c{c}o]{liroy.lourenco@sydney.edu.au}

\author{A.~P. Chippendale}
\affiliation{CSIRO Space and Astronomy, PO Box 76, Epping, New South Wales 1710, Australia}


\newcommand{\achievedSuppresion}{\qty{31}{\decibel}}
\newcommand{\tsysPercent}{\qty{1.5}{\percent}}
\newcommand{\SEFDPercent}{\qty{3.1}{\percent}}
\newcommand{\taperValue}{\qty{-15}{\decibel}}
\newcommand{\holoDifference}{$\pm$\qty{0.4}{\percent}}
\newcommand{\calculatedSupression}{\qty{277}{\decibel}}

\begin{document}

\begin{acronym}[]
    \acrodef{ASKAP}{Australian SKA Pathfinder}
    \acrodef{RFI}{radio frequency interference}
    \acrodef{CSIRO}{Commonwealth Scientific and Industrial Research Organisation}
    \acrodef{MRO}{Murchison Radio-astronomy Observatory}
    \acrodef{FPGA}{field programmable gate array}
    \acrodef{RF}{radio frequency}
    \acrodef{SOI}{signal-of-interest}
    \acrodef{ACM}{Array Covariance Matrix}
    \acrodef{SEFD}{system equivalent flux density}
    \acrodef{CFM}{Conjugate Field Match}
    \acrodef{maxSNR}{maximum Signal-to-noise ratio}
    \acrodef{PAF}{phased array feed}
\end{acronym}

\begin{abstract}
This paper presents the effects of \ac{RFI} mitigation on a radio telescope's sensitivity and beam pattern. It specifically explores the impact of subspace-projection mitigation on the \ac{PAF} beams of the \ac{ASKAP} telescope. The goal is to demonstrate \ac{ASKAP}'s ability to make science observations during active RFI mitigation. The target interfering signal is a self-generated clock signal from the digital receivers of \ac{ASKAP}'s \acp{PAF}. This signal is stationary, so we apply the mitigation projection to the beamformer weights at the beginning of the observation and hold them fixed. We suppressed the unwanted narrowband signal by \achievedSuppresion, to the noise floor of an \qty{880}{\second} integration on one antenna, with a typical degradation in sensitivity of just \tsysPercent. Sensitivity degradation over the whole 36 antenna array of \SEFDPercent \ was then measured via interferometric assessment of \ac{SEFD}. These measurements are in line with theoretical calculation of noise increase using the correlation of the beam weights and \ac{RFI} spatial signature. Further, degradation to the main beam's gain is \holoDifference~ on average at the half-power point, with no significant change to the gain in the first sidelobe and no variation during extended observations; also consistent with our modelling. In summary, we present the first demonstration of mitigation via spatial nulling with \acp{PAF} on a large aperture synthesis array telescope and assess impact on sensitivity and beam shape via \ac{SEFD} and holography measurements. The mitigation introduces smaller changes to sensitivity than intrinsic sensitivity differences between beams, does not preclude high dynamic range imaging and, in continuum \qty{1}{\mega\hertz} mode, recovers an otherwise corrupted holography beam map and usable astronomical source correlations in the RFI-affected channel.
\end{abstract}

\section{Introduction}
\label{sec:int}

\acresetall
For astronomers to adopt active \ac{RFI} mitigation, they need to understand the impact of mitigation on telescope performance and calibration.

Active \ac{RFI} mitigation algorithms can help sensitive radio telescopes operate over scientifically interesting bandwidths in the congested modern radio spectrum. \ac{RFI} continues to increase with population and technology use  \citep{CommitteeonRadioAstronomyFrequencies1997CRAFAstronomy}. Telescopes are also becoming more susceptible to \ac{RFI} as they are built to operate with higher sensitivities and bandwidths. The \ac{ASKAP} telescope \citep{Hotan2021AustralianDescription} used in this work operates over \qtyrange{700}{1800}{\mega\hertz} in frequency bands where satellite navigation, aviation, and terrestrial mobile communication base stations have a primary service allocation \citep{RR2016}. Active mitigation algorithms could enable the use of the \qtyrange{1.15}{1.3}{\giga\hertz} band that is largely avoided by \ac{ASKAP} due to satellite navigation signals. 

Radio astronomers currently mark and discard (flag) \ac{RFI}-affected data so that it does not affect scientific conclusions. Unfortunately, discarding data reduces measurement sensitivity as the discarded data contains valuable information from astronomical sources. In the case of a synthesis imaging array like \ac{ASKAP}, flagging also changes the secondary beam, or point spread function, and thereby image resolution and dynamic range. Although effective when carefully implemented, flagging does not exploit all the information the system provides. In particular, the interferer's spatial information is not fully considered. 

\Acp{PAF} allow for the recovery of spatial information of \ac{RFI} and cosmic sources. Using this spatial information, projection techniques for aperture arrays and \ac{PAF}s create nulls in the direction of the \ac{RFI} \citep{Leshem2000MultichannelAstronomy, Hellbourg2012ObliqueAstronomy, Jeffs2008BiasInterference}, which could reduce the amount of discarded and flagged data. The viability of this technique for the \ac{ASKAP} \ac{PAF} has already been explored in simulation \citep{Black2015Multi-tierTelescope} and demonstrated on six antennas during commissioning \citep{Hellbourg2017SpatialArray}. \cite{Black2015Multi-tierTelescope} showed that mitigation via spatial nulling at the \ac{PAF} beamformer is most effective. The \ac{PAF} primary beam nulls are spatially broader and, therefore, don't have to be updated as rapidly to track moving interferers as would be the case for secondary beam nulls formed at the correlator.

Despite promising progress of active \ac{RFI} mitigation radio in radio astronomy \citep{Fridman2001RFIAstronomy, Series2013TechniquesAstronomy}, much work remains to implement these techniques as part of a holistic strategy in routine, large-scale telescope operations in combination with `traditional' \ac{RFI} mitigation techniques.

This paper begins to address this gap, building on work by \cite{Chippendale2017InterferenceTelescope} in which a narrowband tone generated by the \ac{PAF} digital back-end was suppressed by up to \qty{20}{\decibel} using a modified \ac{ASKAP} Mk.~II \ac{PAF} on the \qty{64}{\metre} Parkes telescope. In this instance, the unwanted signal is static (not moving) relative to the receiver and mitigated at the start of the observation. This static case is an essential first step toward mitigating more pernicious forms of \ac{RFI} from satellite and aviation systems that are dynamic (moving) and need to be mitigated in as close to real-time as possible. Note that in cases where the interference is external to the receiver (or downstream electronics), \ac{RFI} will appear dynamic from the telescope's point of view because the telescope tracks a source on the sky. 

We present simulated forward predictions of the effects of spatial nulling via \ac{PAF} beamforming on astronomical figures of merit, comparing two subspace-projection-based algorithms with attention to the impacts on sensitivity (noise), bandpass (gain), and beam shape, including polarisation response. Section 2 describes the \ac{ASKAP} \ac{PAF}, the signal model and an overview of the spatial filtering algorithms. Section 3 gives simulated forward predictions of the effects of the mitigation. Finally, section 4 presents the implementation and measured performance on \ac{ASKAP}.

\section{Background}
\label{sec:back}
\begin{figure}[!ht]
    \centering
    \includegraphics[width=0.95\textwidth]{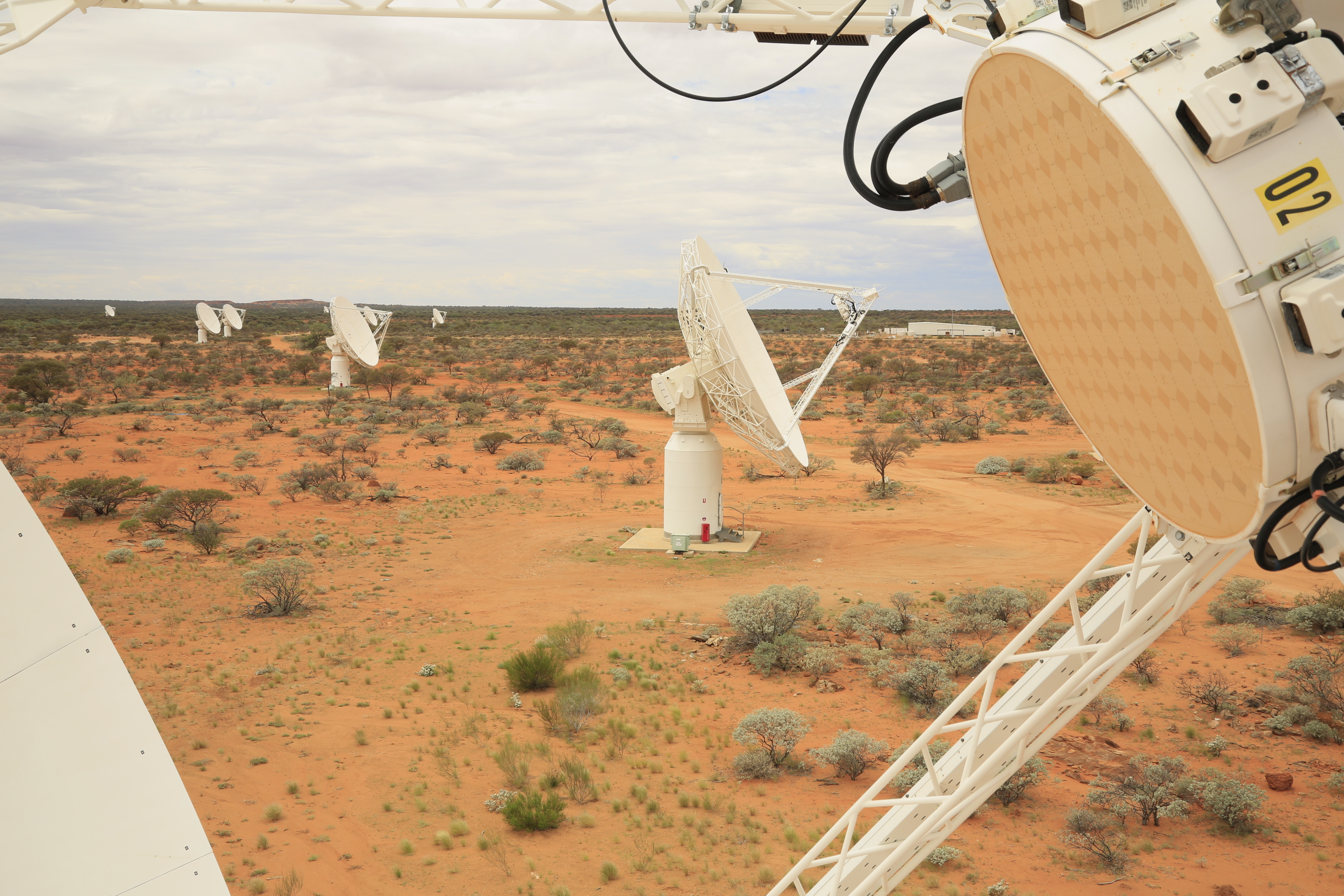} 
    \caption{\qty{12}{\metre} \ac{ASKAP} antennas with \ac{PAF}s at their foci at Inyarrimanha Ilgari Bundara, the \acs{CSIRO} \acl{MRO}. Credit: \acs{CSIRO}}
    \label{fig:askapImage}
\end{figure}

The \acl{ASKAP} telescope comprises thirty-six \qty{12}{\metre}~parabolic reflectors at Inyarrimanha Ilgari Bundara, the \acs{CSIRO} \acl{MRO}.  This unique radio-quiet site is protected by rigorous testing and shielding of observatory electronics and legislative coordination of spectrum usage up to a radius of \qty{260}{\kilo\metre} \citep{Wilson2013MeasuresAustralia}. Figure \ref{fig:askapImage} shows the Mk. II \acf{PAF} at the focus of each antenna. Behind the 188 chequerboard elements (94 per polarisation) is the receiver electronics package housing the low noise amplifiers (per antenna element), three \ac{RF} band selection filters and \ac{RF}-over-Fibre transmitter \citep{Hotan2021AustralianDescription}.

The self-generated \ac{RFI} is generated by a  \qty{256}{\mega\hertz} \ac{FPGA} clock that is multiplied by an oversampling rate of 32/27 to read out the digital receiver’s 1 MHz resolution oversampled coarse filter bank \citep{Tuthill2012DevelopmentASKAP}. Relative delays of this unwanted signal between digital ports differ significantly from relative delays of the desired astronomical signals entering via the reflector and \ac{PAF}. This difference in the spatial signature allows the \ac{RFI} to be cancelled at the beamformer output with little impact on the desired signal. The digital receiver is housed in the stable environment of the \ac{ASKAP}’s central building, so we expect this \ac{RFI} and its path into the signal chain to be stable except for resets of the digital receiver, which do not occur during observing.

The fundamental of this $32 / 27 \times \qty{256}{\mega\hertz} = \qty{303.4}{\mega\hertz}$ clock signal is aliased and appears at the frequency labeled
\[ F_s/2 + (F_s/2 - 303.4) = \qty{976.6}{\mega\hertz} \]
when observing in \ac{ASKAP}'s \qtyrange{700}{1200}{\mega\hertz} band with a sampling rate of $F_s=\qty{1280}{\mega\hertz}$. 

\begin{figure}[!ht]
    \centering
    \includegraphics[width=0.95\textwidth]{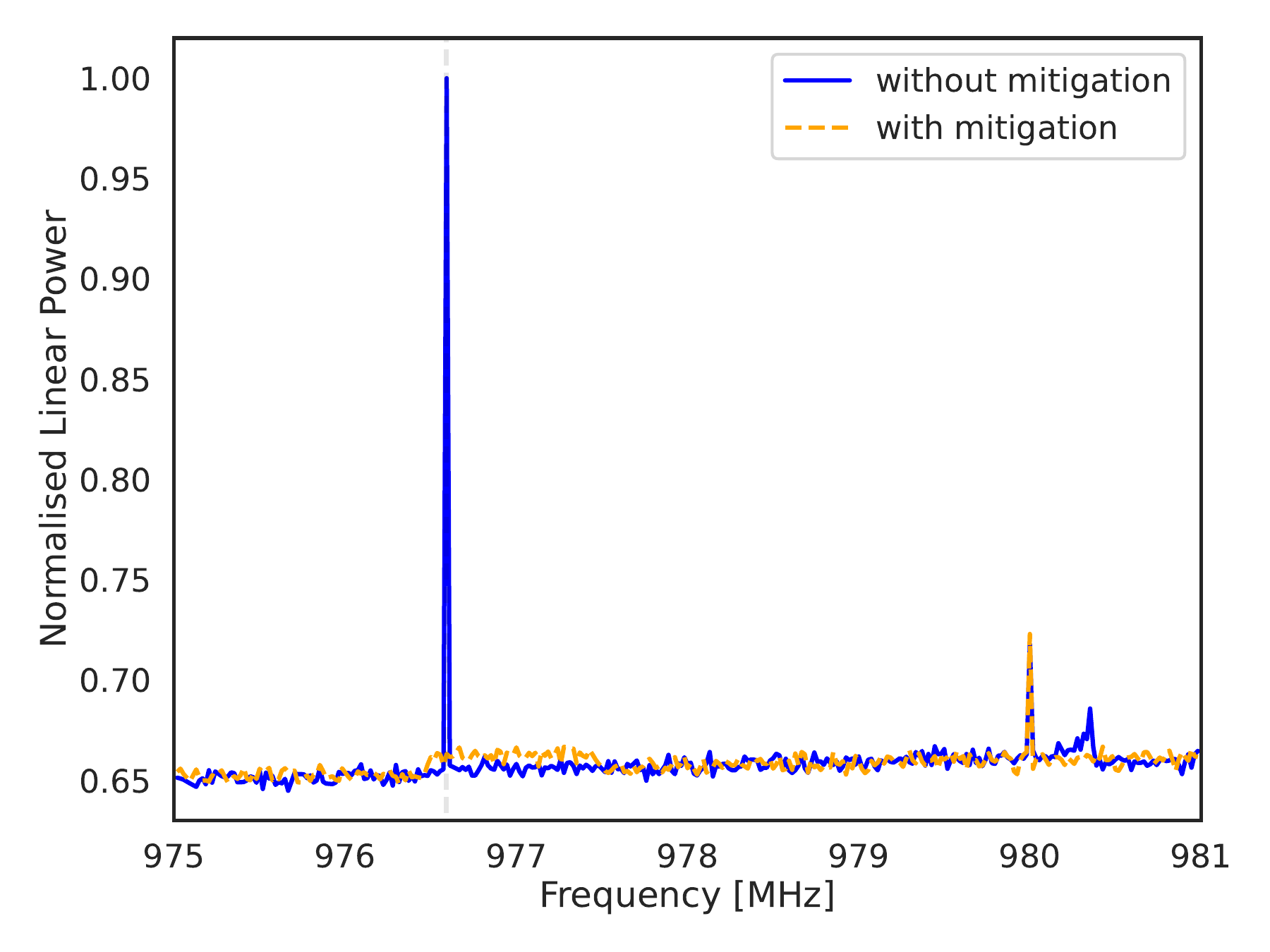}
    \caption{
Beamformed single-antenna power spectra with mitigation (orange) and without (blue). Beamformer weights are fixed over each \qty{1}{\mega\hertz} coarse channel before further channelisation to \qty{18.5}{\kilo\hertz} resolution. The \qty{976.6}{\mega\hertz}  unwanted clock signal is successfully mitigated but with an increase in the system temperature for that coarse channel. The difference between spectra at \qty{980.3}{\mega\hertz} is due to these observations occurring at different times. The mitigation, using the oblique projection, was performed whilst observing a flux reference source, Virgo A, assuming the flux model of \cite{Ott1994AnCalibrators} using a single \ac{ASKAP} antenna.
}
    \label{fig:clocksignal}
\end{figure}

Figure \ref{fig:clocksignal} shows the clock signal in the fine filter bank power spectrum without mitigation (blue) and with mitigation (orange). The mitigation suppresses the unwanted interference, but also increases the system temperature of the \qty{1}{\mega\hertz} channel in which beamformer weights were adjusted to implement the mitigation. 

\subsection{Signal model}
The instantaneous output for a single \ac{ASKAP} \ac{PAF} (Figure \ref{fig:askapImage}) is an $M$-dimensional vector of complex voltages ($M = 188$)
\begin{equation}\mathbf{x}[n] = x_\text{SOI}[n]\textbf{a}_\text{SOI} + x_\text{RFI}[n]\textbf{a}_\text{RFI} + \mathbf{x}_\text{z}[n]\end{equation}
where $x_\text{SOI}[n]$ and $x_\text{RFI}[n]$ are the scalar amplitude time samples of the astronomical \ac{SOI} and \ac{RFI} (unwanted signal to mitigate) respectively. $\textbf{a}_\text{SOI}$ and $\textbf{a}_\text{RFI}$ are the spatial signature vectors of the \ac{SOI} and \ac{RFI}, respectively. $\mathbf{x}_\text{z}[n]$ is the total system noise (the radio sky, feed and receiver). \ac{ASKAP}'s beamformer \citep{Hotan2021AustralianDescription} computes an \ac{ACM}, which is the pairwise correlation of each port with every other port for a single \qty{1}{\mega\hertz} coarse filter bank channel
\begin{equation} \textbf{R} = \langle \textbf{x}[n]\cdot \textbf{x}[n]^H \rangle = \frac{1}{L} \sum_{n=1}^{L} \textbf{x}[n]\textbf{x}^{H}[n]\end{equation}
where $(.)^{H}$ is the Hermitian (complex conjugate) transpose, and $\langle.\rangle$ is the average operator. 
The spatial filtering algorithms used below use this \ac{ACM} to estimate the \ac{RFI} spatial signature ($\textbf{a}_{RFI}$) and subsequently suppress its contribution. For typical moving \ac{RFI} $\textbf{a}_{RFI}$ must be estimated frequently.  For the stationary clock signal being mitigated here, $\textbf{a}_{RFI}$ can be estimated once at the beginning of an observation because it does not change with time or \ac{PAF} motion.

Under typical operation, a set of beamforming steering vectors or ``beamformer weights'' are calculated at the beginning of the observation and loaded to the beamformer to create up to 36 dual-polarized beams on \ac{ASKAP}. The 188 weights correspond to a weighting of each \ac{PAF} port and are calculated per beam, per \qty{1}{\mega\hertz} channel via the \ac{ACM} using a maxSNR algorithm from an on-source and off-source pointing \citep{Hotan2021AustralianDescription}. This beamforming process is most easily understood as applying the weights as a weighted sum of the \ac{PAF} array element voltages
\begin{equation}\mathbf{y}[n] = \mathbf{w}^H \mathbf{x}[n].\end{equation}

The goal is to update these beamformer weights ($\mathbf{w}$) by projecting them onto a subspace orthogonal to the interference subspace, forcing a deep null in the ``direction'' of the interferer but preserving beam gain in the direction of the original beam that was designed to receive the \ac{SOI}.

\subsection{Mitigation algorithm}
Subspace projection refers to a subset of spatial filtering signal processing techniques involving a linear projection to cancel the unwanted signal in the beamformed output of an array. Two projections are considered in this research: orthogonal \citep{Leshem2000Radio-astronomicalInterference} and oblique \citep{Behrens1994SignalOperators, Hellbourg2012ObliqueAstronomy}. \cite{Hellbourg2012ObliqueAstronomy} and \cite{Warnick2018PhasedCommunications} give a full description of both projections.  Both use an eigenvalue decomposition of the \ac{ACM} to determine the `direction' of the \ac{RFI}

\begin{equation} \mathbf{R} = \mathbf{U} \Lambda \mathbf{U}^{H}. \label{eqn:EVD}\end{equation}
Each column of $\mathbf{U}$ is an eigenvector of $\mathbf{R}$, corresponding to the eigenvalues contained along the diagonal of the matrix $\Lambda$ sorted in descending power. $\mathbf{U}$ may be partitioned into columns $\mathbf{U}_{RFI}$ defining the \ac{RFI} subspace and $\mathbf{U}_{SOI + z}$ defining the signal plus noise subspace

\begin{equation}\mathbf{U} = [\mathbf{U}_\text{RFI}|\mathbf{U}_{\text{SOI} + z}].\end{equation}
In the case of the self-generated interference, Figure \ref{fig:EigenvalueSpectra} shows there is only one dominant eigenvalue.  The corresponding eigenvector $\mathbf{a}_\text{RFI}$ comprises a one-dimensional \ac{RFI} subspace $\mathbf{U}_\text{RFI}$.

An orthogonal projection matrix $\mathbf{P}_\text{ortho}$ is defined as follows, where $\textbf{I}$ is the identity matrix 

\begin{equation}
    \label{eqn:ortho}
    \mathbf{P}_\text{ortho}= \mathbf{I} - \mathbf{U}_\text{RFI}\left(\mathbf{U}_\text{RFI}^{H}\mathbf{U}_\text{RFI}\right)^{-1}\mathbf{U}_\text{RFI}^{H}.
\end{equation}
Whilst orthogonal projection mitigates the unwanted signal, it introduces undesirable distortions to the primary beam shape. Therefore, oblique projection is preferred as it better preserves the desired beam shape \citep{Hellbourg2012ObliqueAstronomy}. Using the calculated maxSNR beam weights ($\mathbf{w}$) as our spatial signature $\mathbf{a}_\text{SOI}$, we determine a projection matrix $\mathbf{P}_\text{obliq}$, which not only mitigates the interferer but maintains the steering direction and gain 
\begin{equation}
    \label{eqn:obliq}
    \mathbf{P}_\text{obliq}= \mathbf{a}_\text{SOI}(\mathbf{a}_\text{SOI}^{H}\mathbf{P}_\text{ortho}\mathbf{a}_\text{SOI})^{-1}\mathbf{a}_\text{SOI}^{H}\mathbf{P}_\text{ortho}.
\end{equation}
To obtain updated (``mitigated'') weights, the chosen projection matrix is applied to the original weights 
\begin{equation}
    \mathbf{w}_\text{proj} = \mathbf{P}^{H} \mathbf{w}
\end{equation}
before uploading them to the beamformer where they are used to calculate a beamformed voltage time series with suppressed \ac{RFI}
\begin{equation}
    \mathbf{y}_\text{proj}[n] = \mathbf{w}_\text{proj}^{H}\mathbf{x}[n].
\end{equation}

\section{Simulated forward predictions}
\label{sec:sims}
\begin{figure}[ht]
    \centering
    \includegraphics[width=0.95\textwidth]{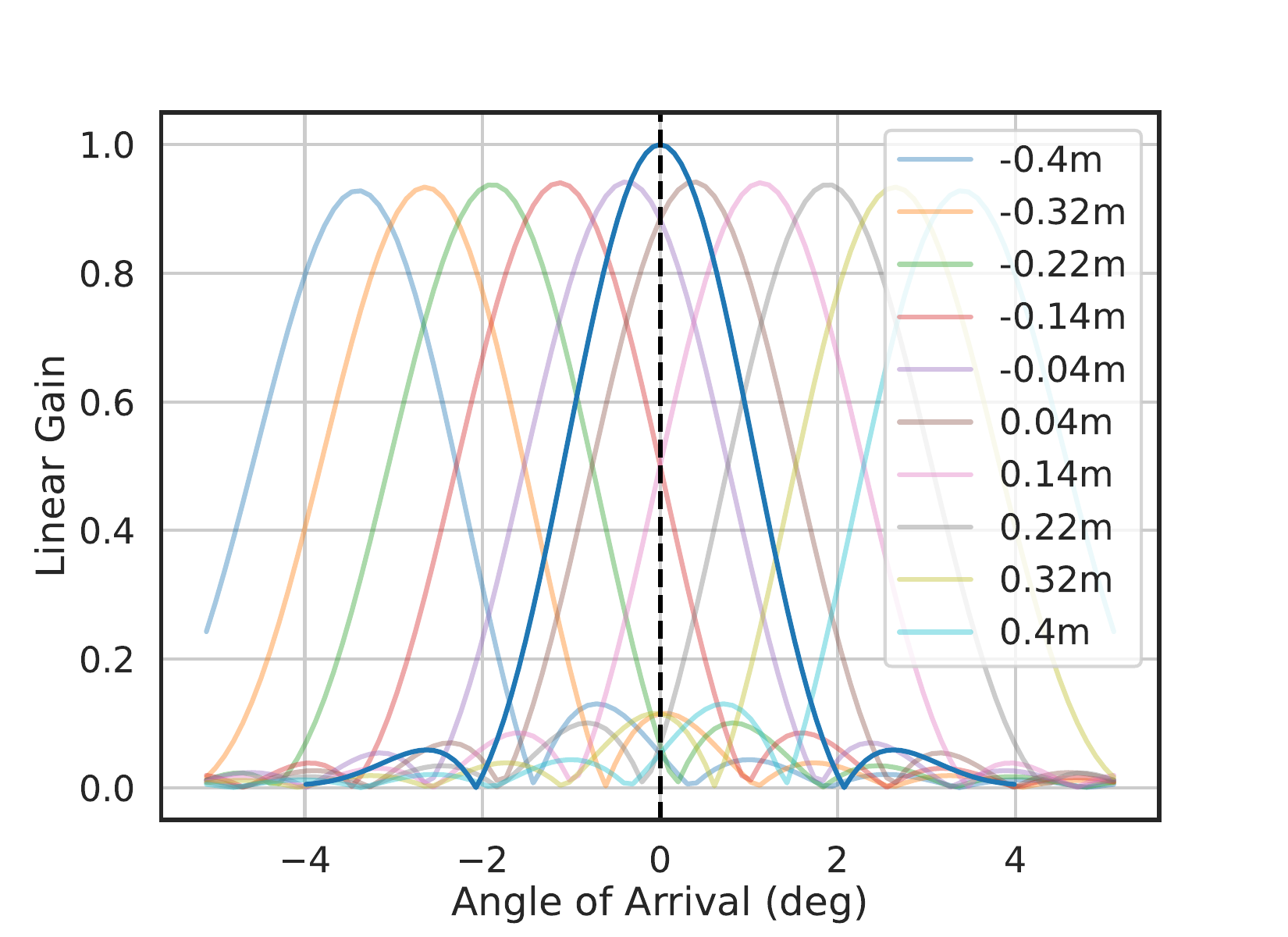}
    \caption{A model was created based on the \ac{PAF} port spacing and diameter of the antenna. The background shows the x-polarisation element response for the centre row of ports using a $\taperValue$ taper. The background curves are weighted and summed to form a beam in the foreground in a process called digital beamforming.}
    \label{fig:xpolElementResponse}
\end{figure}

The voltage pattern of individual \ac{ASKAP} \ac{PAF} elements is estimated based on the geometric optics analysis of \cite{Baars2007TheCommunication} for the illumination of a parabolidal reflector by laterally offset feeds
\begin{IEEEeqnarray*}{rl}
    A_k(l,m)=\sqrt{2.56} \int_{0}^{1}2rF(r)J_{0}\Bigl[ \Bigr. r\Bigl\{ \Bigr.& (l - sx_{k})^2 \IEEEyesnumber \\
                                                &  \left. \left.+ (m - sy_{k})^2 \right\}^{0.5}\right]dr
\end{IEEEeqnarray*}
where J\textsubscript{0} is the Bessel function of the first kind and order zero. The illumination function $F(r)= 1 - (1 - \tau) r^2$
based on the geometry of the reflector and \ac{PAF} element positions (above). A one-dimensional slice showing the x-pol element response for the centre row (ten ports) of the \ac{ASKAP} \ac{PAF} is plotted in the background of Figure \ref{fig:xpolElementResponse}. 

To create the beam ($\mathbf{y}[n]$) at the boresight in the foreground of Figure \ref{fig:xpolElementResponse}, the model uses maximum gain weights that effect a \ac{CFM}; that is, the intersections of the background curves with the dashed vertical line are the weights $\mathbf{w}$ used to steer the beam. The exact process is repeated for the y-polarisation ports but is not shown here. 

\ac{PAF} ports are weighted by a steering vector and summed to create beams. The beam in the foreground of Figure \ref{fig:xpolElementResponse} shows a high-gain main lobe centred about the dotted line and terminates on either side with a null. The first and second side lobes are also discernible, separated by nulls. The model evaluates the effects of the orthogonal and oblique projection on six equally spaced beams ($\approx0.9^\circ$ degrees apart) across the \ac{PAF} for both polarisations. This beam separation was chosen based on the close-pack square $6\times6$ beam footprint used by \ac{ASKAP} \citep{Hotan2021AustralianDescription}. 

\begin{figure}[ht]
    \centering
    \includegraphics[width=1\textwidth]{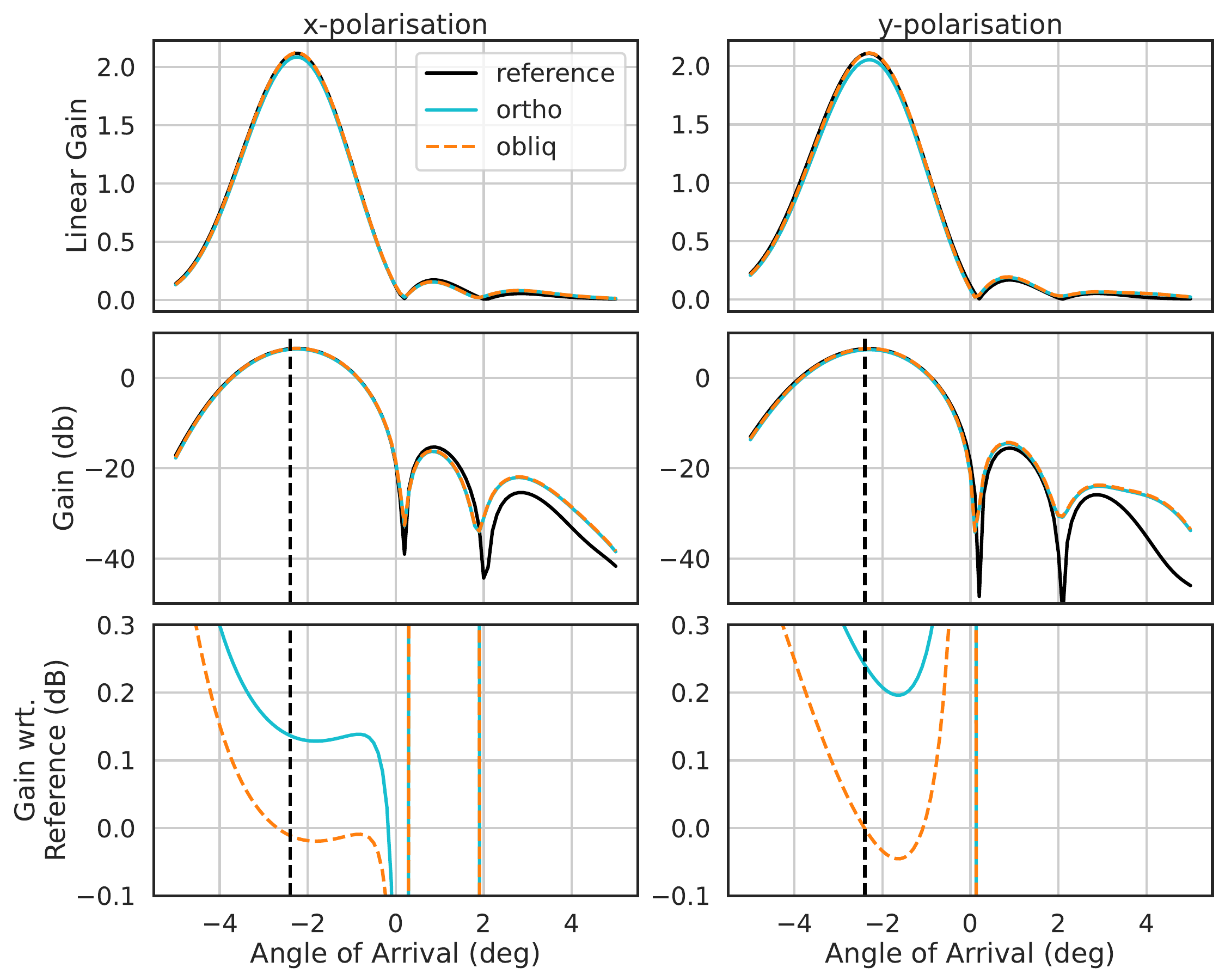}
    \caption{X and Y polarisation beamformed response in the direction -\qty{2.4}{\degree}, before and after mitigation. The rows show the linear gain, gain in decibels and gain in decibels with respect to the reference (unmitigated) beam.  This result is for the case where the \ac{RFI} enters each \ac{PAF} port with a measured \ac{RFI} spatial signature. Variations to the beam shape are reduced in the main lobe when using oblique projection. In both cases, introduced variations to suppress the unwanted signal are confined to the outer lobes.}
    \label{fig:beamshapeComparison}
\end{figure}

The effects of mitigation, using an estimated 1-D \ac{RFI} spatial signature of the stationary internal clock signal determined from an \ac{ASKAP} \ac{ACM},  were assessed on changes to model beam shape (Figure \ref{fig:beamshapeComparison}) and suppression calculated as follows
\begin{equation}
    20\log_{10}
\frac{\lvert \mathbf{w}^H \mathbf{U}_\text{RFI} \rvert}{\lvert \mathbf{w}_\text{proj}^{H} \mathbf{U}_\text{RFI} \rvert} \ \text{dB}.
\label{eqn:suppression}
\end{equation}
As expected from previous experiments \citep{Chippendale2017InterferenceTelescope}, the projection algorithms suppress the unwanted signal.
The calculated predicted suppression of the unwanted signal across the six simulated beams yields an upper limit on the suppression greater than \qty{250}{\decibel}. In other words, how orthogonal the \ac{RFI} spatial signature can be made given, for example,  the numeric precision of the calculations. In practice, this will be limited by the noise floor and the degree to which an accurate \ac{RFI} spatial signature vector can be estimated.

Figure \ref{fig:beamshapeComparison} shows the X and Y polarisation beamformed response in the direction -\qty{2.4}{\degree} (black vertical dashed line), before and after mitigation using a measured \ac{RFI} spatial signature. The rows show the linear gain, gain in decibels and gain (in decibels) with respect to the reference (unmitigated) to highlight differences in maximum gain of the main beam.

After mitigation, variations to the beam patterns were observed mostly in the outer side lobes (starting at the second null). The bottom panel of Figure \ref{fig:beamshapeComparison} shows differences in the main beam are reduced when using oblique projection. Two instances of the self-generated interferer \(\mathbf{a}_\text{RFI}\) were considered in the model (\ref{appendix::coherent_vs_incoherent}); a uniform amplitude arriving at all ports with and without randomised delay/phase.  That is all ports with different delay (randomised) and all ports with the same (aligned) phase, respectively. It was found that distortions to the beam were reduced when the phase of the generated signal was randomised. The estimated \ac{RFI} spatial signature in Figure \ref{fig:beamshapeComparison} is consistent with a clock signal arriving at the \ac{PAF} ports with randomised phase due to differing delays in the signal path of each port. Using this property, we can start to preselect \ac{RFI}, based on its phase as being a suitable candidate for subspace projection \ac{RFI} mitigation. Similarly, for a one-dimensional RFI subspace $\mathbf{u}_\text{RFI}$  the correlation $\rho$, between the estimated \ac{RFI} and beam weights, given by their dot product
\begin{equation}
    \rho = \frac{\left| \mathbf{w}^H\mathbf{u}_\text{RFI} \right|}{ \norm{\mathbf{w}} \norm{\mathbf{u}_\text{RFI}}}
\label{eqn:correlationCoefficent}
\end{equation}
can be used to estimate the performance impact of the orthogonal and oblique projections as given by \cite{Hellbourg2012ObliqueAstronomy}. For oblique projection the increase in the noise power is 

\begin{equation}
    P_{noise} = \sigma^{2}_{n} \left( 1 - \left| \rho \right|^{2} \right)^{-1}
\label{eqn:predicted_noise_oblique}
\end{equation}
where $\sigma^{2}_{n}$ is the true noise power.

Simulating changes in T\textsubscript{sys} requires a full simulation of the noise and mutual coupling. Without a full noise model, this paper uses \ac{CFM} (maximum gain) weights in place of the \ac{maxSNR} weights used in operational \ac{ASKAP} beamforming. We, therefore, measure the mitigation's effects on sensitivity using on and off-source measurements from the telescope directly. 

\section{Measured performance on ASKAP}
\label{sec:res}

\begin{figure*}[!ht]
\begin{subfigure}[t]{.475\textwidth}
    \includegraphics[width=0.95\textwidth]{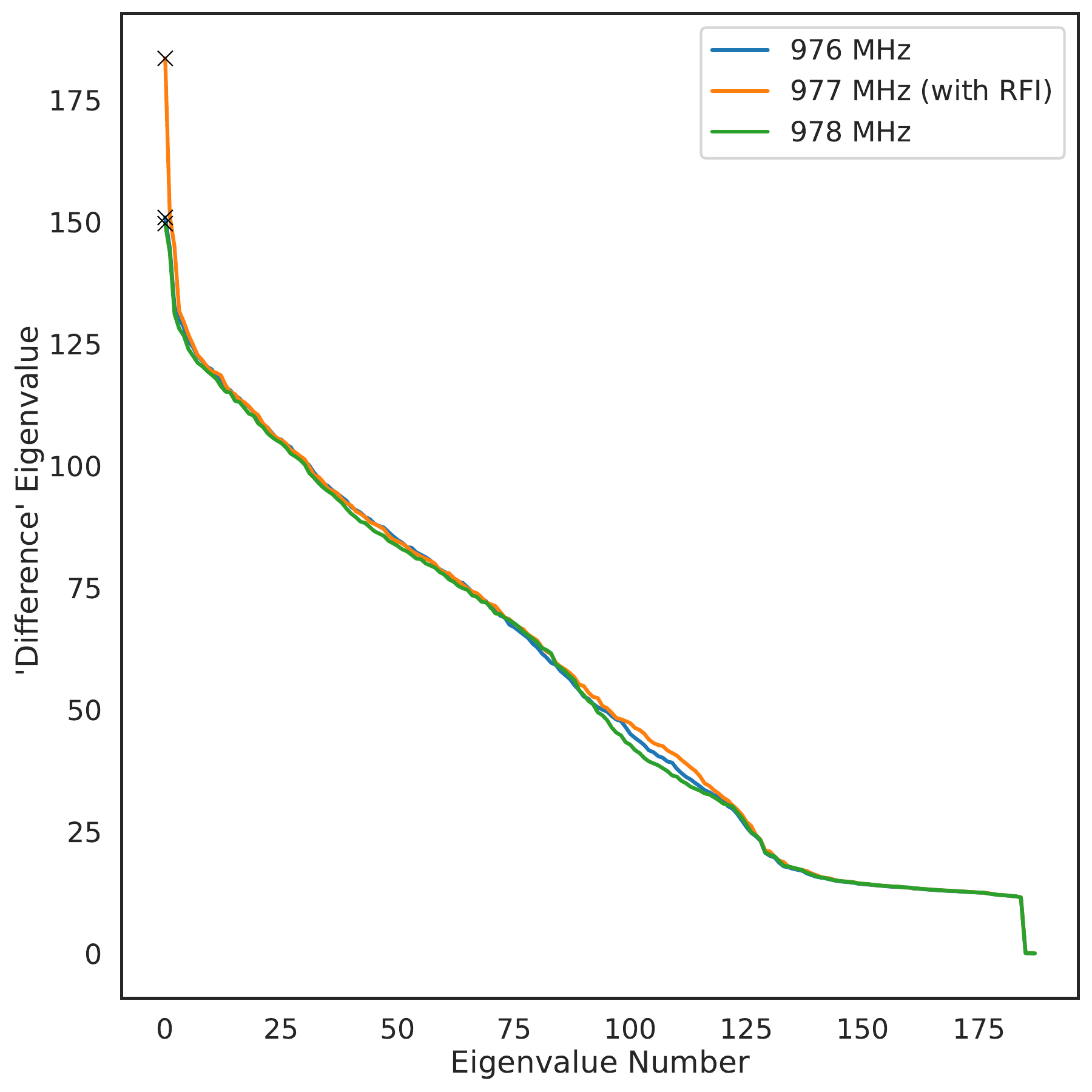} 
    \caption{The eigenvalue spectra of three adjacent coarse \ac{ACM} channels, the centre channel (\qty{977}{\mega\hertz}), contains the self-generated interference. The difference in the first value in the \qty{977}{\mega\hertz} channel is due to the interfering clock signal.}
    \label{fig:EigenvalueSpectra}
\end{subfigure}
\hfill
\begin{subfigure}[t]{.475\textwidth}
    \includegraphics[width=0.95\textwidth]{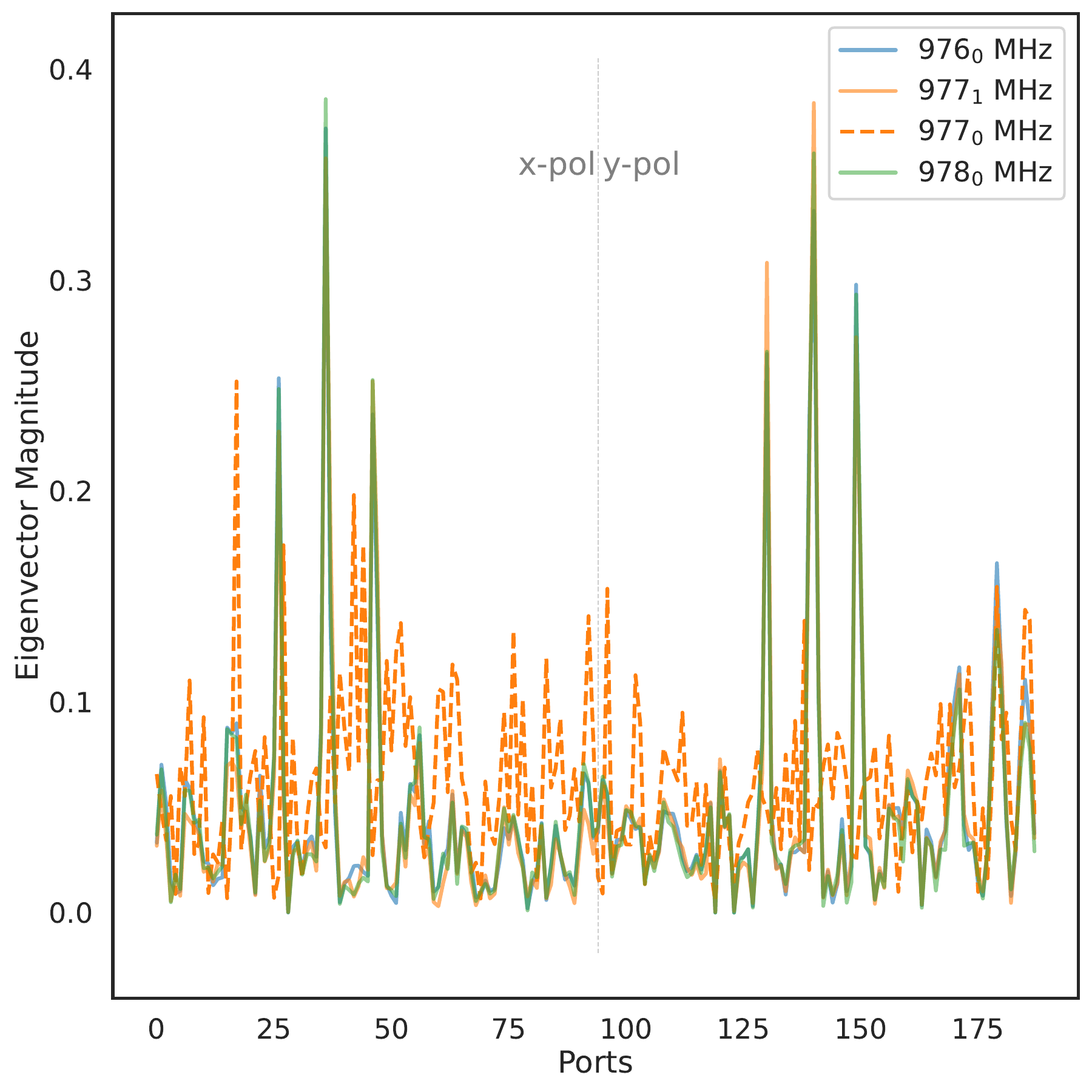} 
    \caption{The dominant eigenvectors of the three adjacent channels, as well as the second eigenvector of the centre channel (with \ac{RFI}), $\mathbf{U}_{{977_1}}$ (the dashed curve). Note how the ${\mathbf{U}_{{976}_0}}$, $\mathbf{U}_{977_{1}}$ and $\mathbf{U}_{978_{0}}$ have a similar structure compared to $\mathbf{U}_{977_{0}}$ (the dashed curve meaning they have the similar weights of ports on the \ac{PAF}).}
    \label{fig:eigenvectors}
\end{subfigure}
\caption{The eigenvalue decomposition of three adjacent \acp{ACM} centred on the \ac{RFI}-affected channel showing the power and `direction' of the eigenmodes.} 
\label{fig:4-label}
\end{figure*}

We conducted three experiments to assess the effects of spatial nulling via \ac{PAF} beamforming on astronomical figures of merit. First, we conducted offline mitigation and calculated the suppression and sensitivity. The second experiment was conducted in real-time using on and off-source measurements whilst observing Virgo A using a single \ac{ASKAP} antenna with and then without mitigation. We measured suppression limited by the noise floor and a beam equivalent system-temperature-on-efficiency $T_\text{sys}/\eta$ across eight beams via the Y-factor ratio of beamformed power between on and off-source pointings. 

Finally, we performed simultaneous measurements across the entire array, using pairs of beams in the same direction. Each beam pair comprises one beam with oblique projection and the other using maxSNR weights. We simultaneously measure the \ac{SEFD} and perform holography on both mitigated and unmitigated beams to isolate changes due to the mitigation from  other variations. This final experiment uses all antennas and a set of 36 beams to quantify the impact on telescope performance of mitigating the narrowband self-generated signal during operational observing modes.

\subsection{Implementation}
Eigenvalue decomposition of the \ac{ACM} is performed to isolate the one-dimensional spatial signature of the self-generated interferer via the dominant eigenvalue. \ac{RFI} moving relative to the telescope would have a higher dimensional subspace \cite[Lourenço \& Chippendale in prep]{Hellbourg2015SUBSPACETELESCOPES}. Looking at the eigenvalue spectra (Figure \ref{fig:EigenvalueSpectra}) of three adjacent channels, we can see that when the self-generated \ac{RFI} is powerful enough, the dominant eigenvalue of the centre channel is easily identifiable --- the max value of the \qty{977}{\mega\hertz} (orange) curve is higher --- compared to the adjacent coarse \qty{1}{\mega\hertz} channels.

When comparing the dominant eigenvectors (Figure \ref{fig:eigenvectors}) of the same three adjacent channels (subscript $[.]_{0}$), as well as the second eigenvector (subscript $[.]_{1}$) of the centre \qty{977}{\mega\hertz} channel (with \ac{RFI}), we see that $\mathbf{U}_{976_{0}}$, $\mathbf{U}_{977_{1}}$ and $\mathbf{U}_{978_{0}}$ have a similar structure compared to $\mathbf{U}_{977_{0}}$ (the dashed curve). $\mathbf{U}_{976_{0}}$, the dashed curve, is our 1-D \ac{RFI} spatial signature ($\mathbf{a}_\text{RFI}$). The other curves are representative of our \ac{SOI} spatial signature ($\mathbf{a}_\text{SOI})$. This is only the case here because the \ac{ACM} is an on-source beamforming \ac{ACM}, so the Sun is in the field-of-view, which gives the structure in the solid curves, dominated by six highly weighted ports (i.e. the ports onto which the Sun's energy is focused by the reflector).

\ac{ASKAP} calculates a weights steering vector per polarisation. Implementation of these algorithms, therefore, requires we calculate  $\mathbf{a_x}_\text{RFI}$ and $\mathbf{a_y}_\text{RFI}$ separately to be consistent with \ac{ASKAP}s existing codebase. We isolate two sub-\acp{ACM} of size $94\times94$ along the diagonal and perform the eigenvalue decomposition on those matrices to obtain $\mathbf{a_x}_{\text{RFI}}$ (first sub-\acp{ACM}) and $\mathbf{a_y}_{\text{RFI}}$ (second sub-\acp{ACM}). That is quadrants I and III of the \ac{ACM} moving clockwise from the top left. In other words the first 94 ports of the \ac{ASKAP} \ac{PAF} (and weights) correspond to the x-polarisation and ports 95 through 188 correspond to the y-polarisation (as illustrated in Figure \ref{fig:eigenvectors}).

Furthermore, \ac{ASKAP} varies the number of ports used in beamforming. Typically for thirty-six beams, only sixty ports per \ac{ASKAP} antenna are used. The number of ports used increases as the number of beams decreases \citep{Hotan2021AustralianDescription}. We compared the effects of using all ports vs sixty ports in simulation and found the results to be comparable.

Finally, in some cases, when the interference-to-noise power is low, the difference observed in Figure \ref{fig:EigenvalueSpectra} is not substantial enough to provide an accurate estimation of $\mathbf{a}_{RFI}$ due to the low power nature of the signal. To overcome this, the \acp{ACM} of the adjacent channels were averaged and subtracted from the centre channel to isolate the spatial signature more easily.  The \ac{ACM}, $\mathbf{R}$ in the eigenvalue decomposition in Equation \ref{eqn:EVD} 

\begin{equation}
    \mathbf{R}^\prime = \mathbf{R}_k - \left[\frac{\mathbf{R}_{k-1} + \mathbf{R}_{k+1}}{2}\right]
\label{eqn:lambda_subtraction}
\end{equation}
is modified to estimate the \ac{RFI} spatial signature.

\begin{figure}[h]
    \centering
    \includegraphics[width=0.95\textwidth]{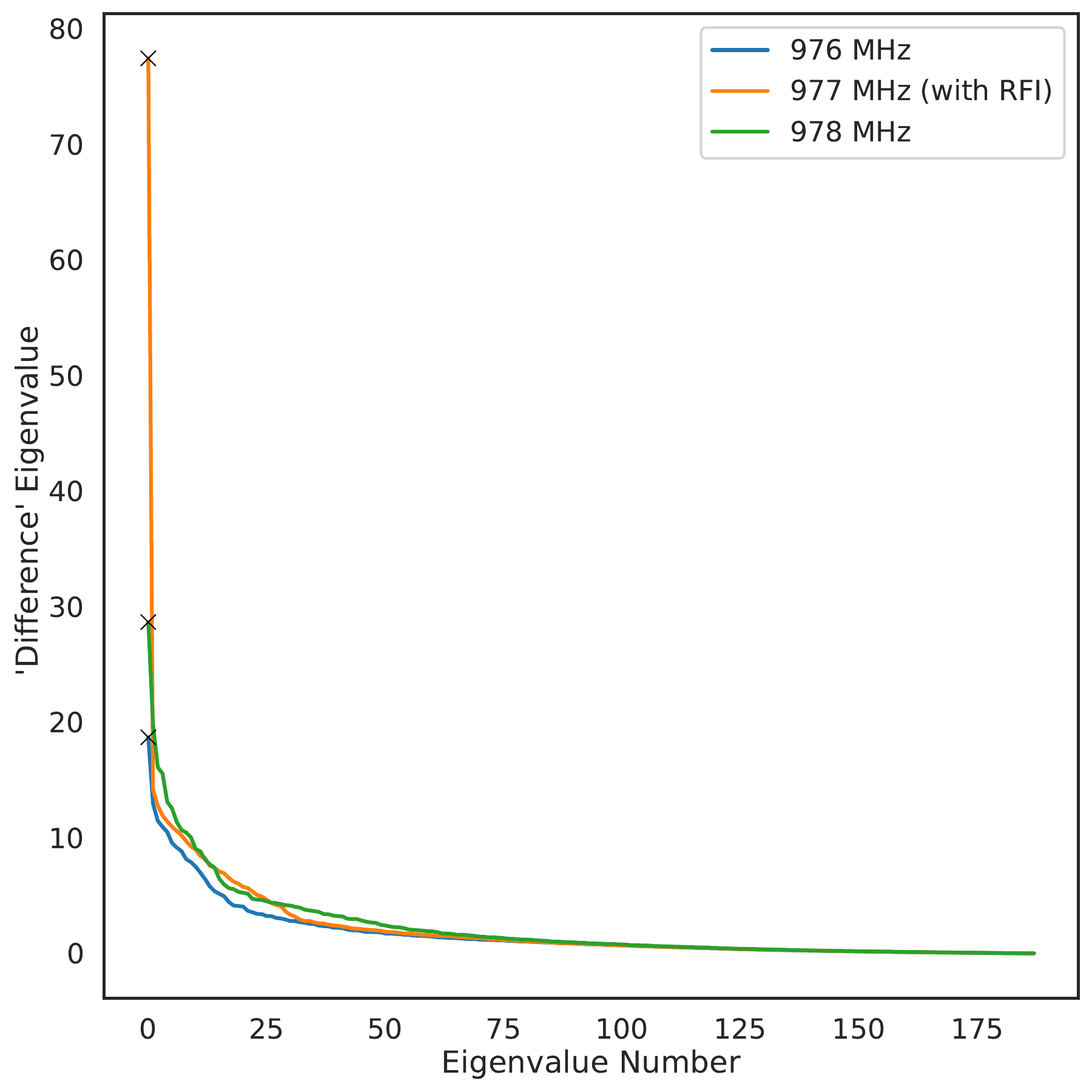} 
    \caption{When the interference-to-noise power is low, by subtracting the mean of the adjacent channels it is easier to isolate an \ac{RFI}- affected channel in the eigenvalue spectra.}
    \label{fig:EigenvalueSpectra_Difference}
\end{figure}

This would not be necessary for typical \ac{RFI} which is much stronger than our test signal, however a useful result nevertheless for identifying and mitigating weak sources of \ac{RFI}. Comparing both methods for calculating $\mathbf{a}_{RFI}$ yielded similar (and parallel determined by the dot product) vectors.

\subsection{Impact on suppression, sensitivity and beam shape}
\begin{figure}[!h]
    \centering
    \includegraphics[width=0.9\textwidth]{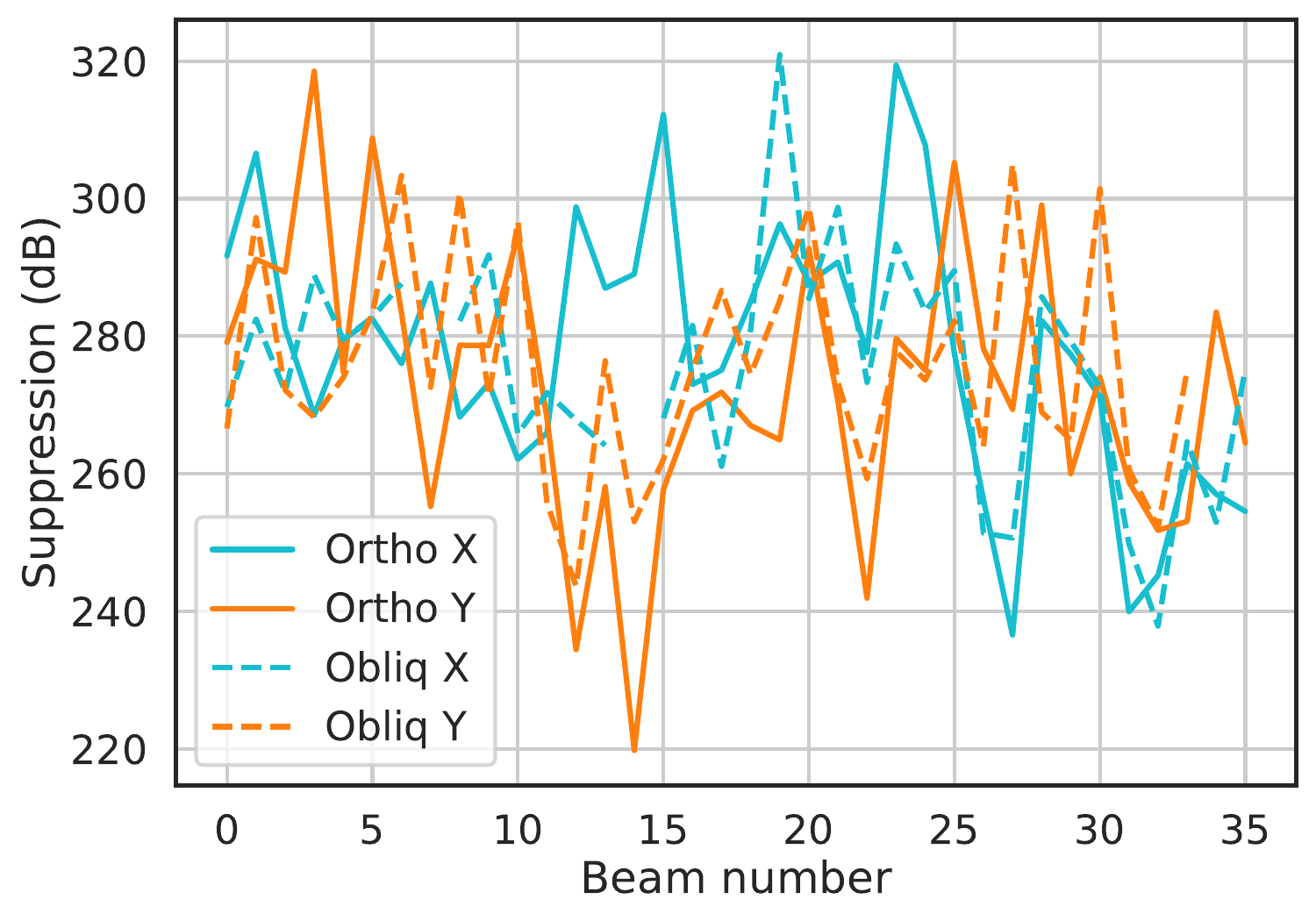} 
    \caption{Calculated suppression of the unwanted signal across beams is approximately \calculatedSupression.}
    \label{fig:supressionExperiment}
\end{figure}

In the first experiment, the mitigation is performed offline. We use \acp{ACM} (containing the unwanted signal) and beamformed \ac{maxSNR} weights from one \ac{ASKAP} antenna (ak18) to determine the calculated suppression across beams. Figure \ref{fig:supressionExperiment} shows a maximum expected suppression of approximately \calculatedSupression~ across all thirty-six beams, consistent with the model.

Using on and off-source \acp{ACM} the Y-factor (on-dish noise performance) can additionally be calculated \citep{Chippendale2014MeasuringCSIRO}: 

\begin{equation}
    Y = \frac{\textbf{w}^H\textbf{R}_{on-src}\textbf{w}}{\textbf{w}^H\textbf{R}_{off-src}\textbf{w}}
\end{equation}
That is, the power ratio between measurements of a known astronomical source and nearby empty sky, respectively. The off-source measurements are routinely obtained at the beginning of a beamformer weight calibration observation and stored in the resulting \ac{ACM} file. Across beams, similar results in terms of suppression were obtained using both projections, as expected. Using offline mitigation we measured a maximum increase to the Y-factor of \qty{17}{\percent} and \qty{15}{\percent} and an average increase of \qty{3}{\percent} and \qty{5}{\percent} for the x and y polarisations, respectively. Up to \qty{6}{\percent} increase could be expected from the calculated value using Equation \ref{eqn:predicted_noise_oblique}. Currently, on ASKAP, as the number of beams increases to the complete set of 36 dual-polarisation beams, the maximum number of ports used to form each beam reduces from 192 to 60 \citep{Hotan2021AustralianDescription}. We assessed the expected increase in noise power using only 60 ports in the above calculation, and recall are only using the sub-ACMs to estimate $\mathbf{a_x}_\text{RFI}$ and $\mathbf{a_y}_\text{RFI}$. We use the indices of the same 60 ports from the beam weights to mask the \ac{RFI} spatial signatures and then calculate the correlation coefficient. 

However, the expected noise power increase is much smaller when calculating the correlation coefficient with an \ac{RFI} spatial signature estimated from the full ACM and using all ports with the x and y maxSNR weights, \qty{0.2}{\percent} and \qty{0.4}{\percent}, respectively, on average. This result shows that the mitigation works much better for a beamformer that can work over all \ac{PAF} ports of all polarisations and suggests how we could improve the mitigation on \ac{ASKAP} in the future. 

The second experiment conducted on 20 June 2022 captures on and off-source measurements using one antenna (ak18 again). The mitigation, using the oblique projection, was performed whilst observing a flux reference, Virgo A, assuming the flux model of \cite{Ott1994AnCalibrators}. Eight beams were formed. First, the antenna was pointed off-source \qty{-9}{\degree} offset in declination, each beam was then subsequently directed toward Virgo A in turn, and fine filter bank data were recorded at \qty{18.5}{\kilo\hertz} resolution. We repeated the experiment with mitigation turned off for comparison. The sensitivity can be determined via the Y-Factor by calculating the beam equivalent system-temperature-on-efficiency as in \cite{Chippendale2016TestingTelescope} and \cite{Chippendale2015MeasuredAntenna,Hayman2010ExperimentalRadiotelescope}.

\begin{equation}
    \frac{T_\text{sys}}{\eta} = \frac{A~S}{2k_B(Y-1)}
    \label{eqn:tsys}
\end{equation}
where A is the area of the dish, S is the known flux of the source, and $k_{B}$ is Boltzmann's constant.

\begin{figure}[!h]
    \centering
    \includegraphics[width=0.90\textwidth]{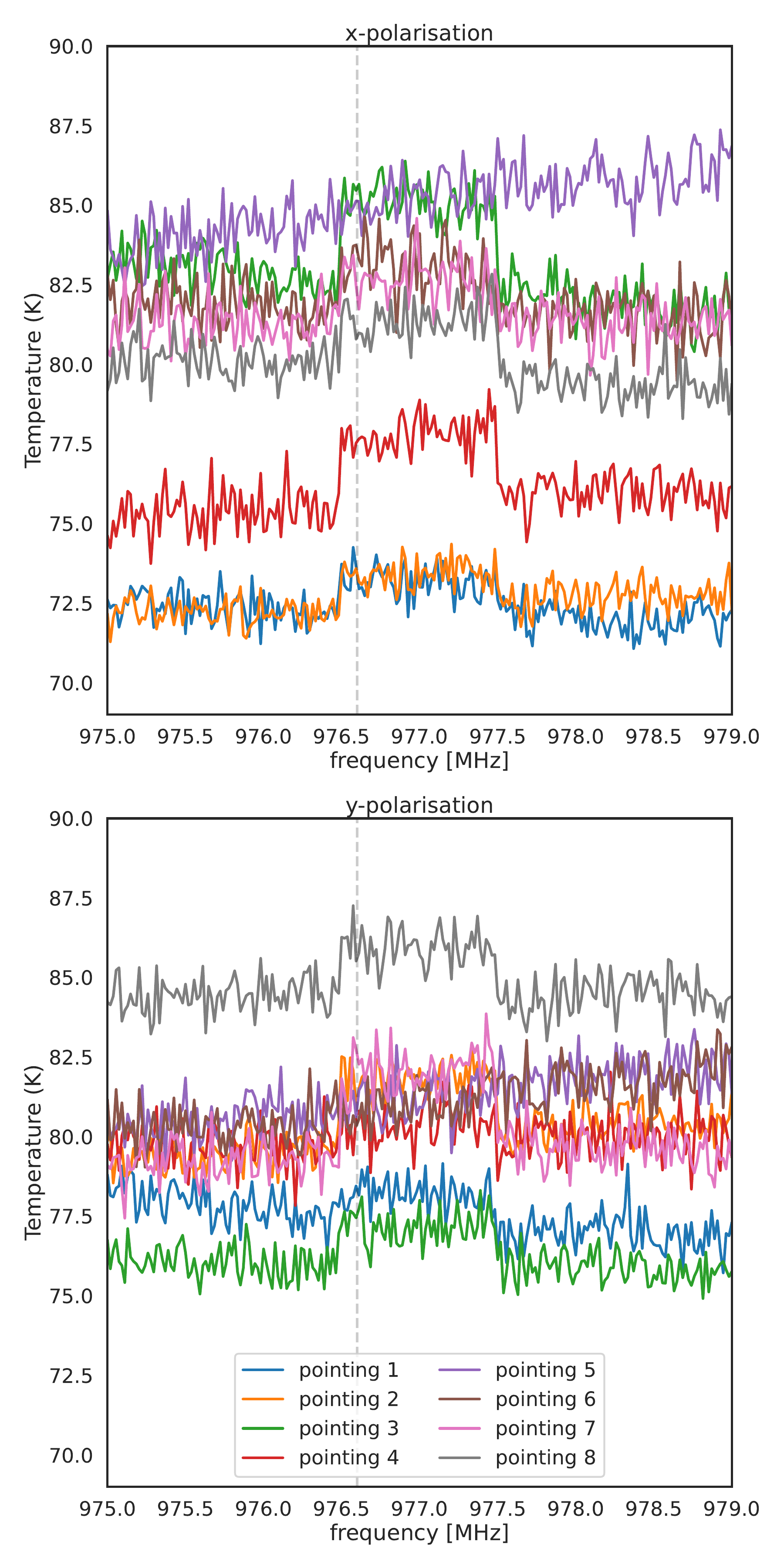}
    \caption{Sensitivity (System-temperature-over-efficiency) with mitigation, shows x-polarization on the top and y-polarization on the bottom across eight different beams. $T_\text{sys}/\eta$ increases at the channel updated using oblique projection compared to maximum signal-to-noise ratio weights (seen in adjacent channels). Beamforming and sensitivity measurements were made using Virgo A and an off-source measurement \qty{-9}{\degree} offset in declination.}
    \label{fig:tsys}
\end{figure}

Figure \ref{fig:tsys} shows the $T_\text{sys}/\eta$ of eight x and y polarization beams across an approximately \qty{3}{\mega\hertz} frequency range centred at the \qty{1}{\mega\hertz} channel containing the unwanted signal. The dashed vertical line demarcates the frequency of the mitigated interfering clock signal. There is a mean $T_\text{sys}/\eta$ of 79.2$\pm$\qty{3.7}{\kelvin},~ 80.6$\pm$\qty{3.8}{\kelvin} and ~79.5$\pm$\qty{3.}{\kelvin} for the ~\qty{976}{\mega\hertz}, ~\qty{977}{\mega\hertz} and ~\qty{978}{\mega\hertz} channels respectively across all beams. All the mitigated channels have a higher $T_\text{sys}/\eta$, except beam 5, due to the upward slope in the curve. In the \qty{1}{\mega\hertz} channel containing the unwanted signal, across all beams, there is a maximum increase of \qty{3.31}{\percent} or \qty{2.66}{\kelvin} in $T_\text{sys}/\eta$ and a mean increase of \qty{1.59}{\percent} or \qty{1.26}{\kelvin}. Note that these increases are smaller than the differences between beams. A mean increase of \qty{1}{\percent} based on the correlations was expected (Equation \ref{eqn:predicted_noise_oblique}). The rest of the band is consistent with \cite{Chippendale2015MeasuredAntenna}. One way to understand the increase in $T_\text{sys}/\eta$ is that the original weights are calculated to yield a maximum signal-to-noise ratio (the lowest $T_\text{sys}/\eta$); any deviation from those weights will by definition be less than the maximum.

Suppression using Equation \ref{eqn:suppression} is not limited by the noise floor. From this experiment, we were able to determine a noise floor limited suppression of \achievedSuppresion, see Figure \ref{fig:clocksignal}. Importantly, we have regained a channel that is otherwise always flagged \citep{Lourenco2024SurveyStatistics} and ultimately discarded.

Recall we are applying the mitigation once at the start of an observation because, in this case, the  1-D \ac{RFI} spatial signature ($\mathbf{u}$) is unchanging during an observation. The similarity of the two signatures \(\Re\left\{\mathbf{u}_{RFI}[t1]^H \mathbf{u}_{RFI}[t2]\right\}>= 0.99\), i.e. they are in the same `direction' over a \qty{7}{\minute} measurement.

\subsection{Concurrent measurements with mitigation on and off}

Two experiments were conducted in January 2023 to compare the effects of the projection techniques simultaneously. First measuring sensitivity across the array, we carried out a \acf{SEFD} measurement whilst observing a flux reference, PKS B1934\textendash638. Then we made a holography measurement to determine changes to the beam shape. 

\begin{figure}[h]
    \centering
    \includegraphics[width=0.90\textwidth]{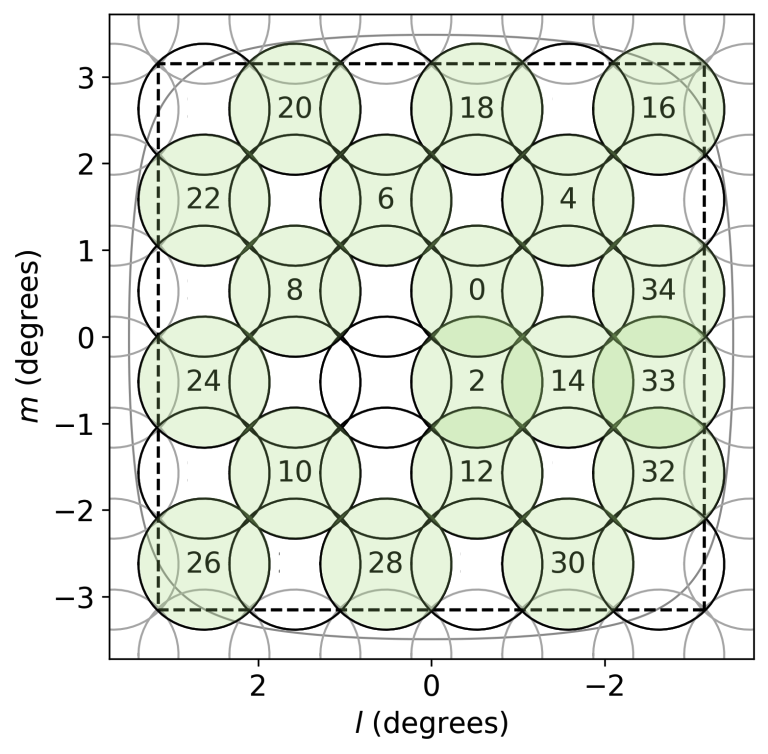}
    \caption{To simultaneously compare the effects of the projection techniques on sensitivity and beam shape, experiments were conducted using a modified close-pack square $6\times6$ footprint. The modified footprint consists of two beams created at each odd position (shaded in green), the first without mitigation and the second with mitigation. Modified from Figure 2 in \cite{McConnell2020TheResults}.}
    \label{fig:footprint}
\end{figure}

A modified \ac{ASKAP} 6$\times$6 square close-pack beam footprint was used in both experiments. The modified footprint creates pairs of beams in the even positions (highlighted in green) of the footprint shown in Figure \ref{fig:footprint}. Thirty-six beams grouped into 18 pairs were created; odd-numbered beams used unmitigated weights, and even beams used updated weights with mitigation using oblique projection. In other words, for every beam in Figure \ref{fig:footprint}, marked in green, there is a mitigated copy in the same position labelled with the subsequent number. Simultaneously measuring beams with and without mitigation isolates changes in the \ac{SEFD} and holography to the mitigation itself.

\subsubsection*{Sensitivity across the Array}

The \ac{SEFD}, measured in Janskys (Jy), is given by \citep{Hotan2021AustralianDescription}

\begin{equation}
    SEFD = \frac{2k_BT_\text{sys}}{A\eta}
    \label{eqn:SEFD}
\end{equation}
and used to characterise the sensitivity of an antenna (and receiver). We use a modification of \ac{ASKAP} standard \ac{SEFD} which uses the correlation coefficient \citep{Perley2008EVLASensitivity} to reduce measurement noise. \ac{SEFD} measurements are a good way of comparing sensitivity between telescopes because it takes into account both the receiver system temperature ($T_\text{sys}$) and the area (A). Note the $T_\text{sys}/\eta$ term, Equation \ref{eqn:tsys}, from the previous experiment. A lower \ac{SEFD} value corresponds to an increased sensitivity. 

\begin{figure}[h]
    \centering
    \includegraphics[width=\textwidth]{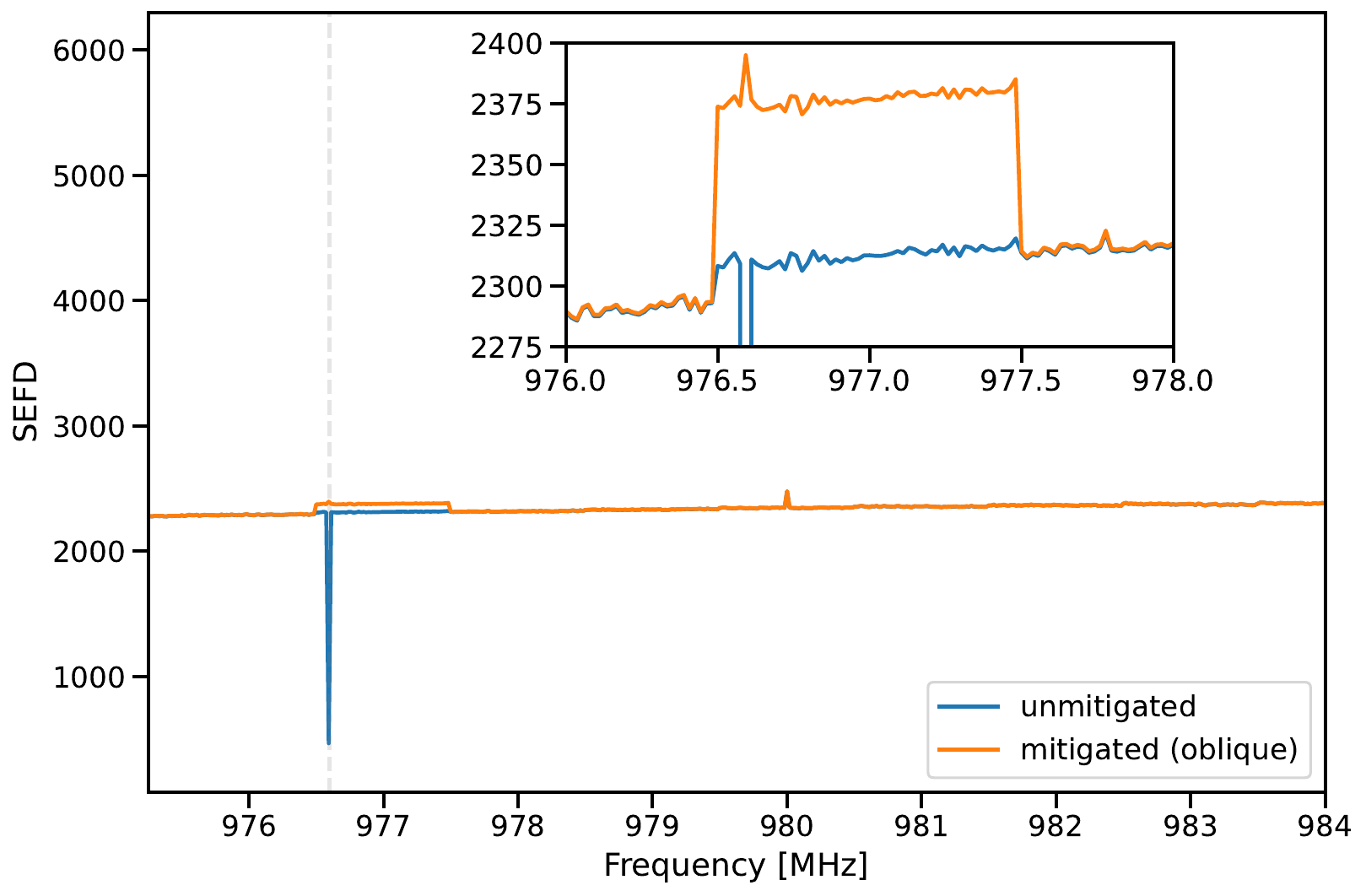}
    \caption{\ac{SEFD} measurement showing unwanted clock signal with and without mitigation (orange and blue, respectively). Similar to a single antenna across the array, mitigation introduces a reduction in sensitivity to the \qty{1}{\mega\hertz} channel. This plot is averaged over all beams and all antennas.}
    \label{fig:SEFD_averaged}
\end{figure}

\begin{figure*}[t]
    \centering
    \includegraphics[width=1\textwidth]{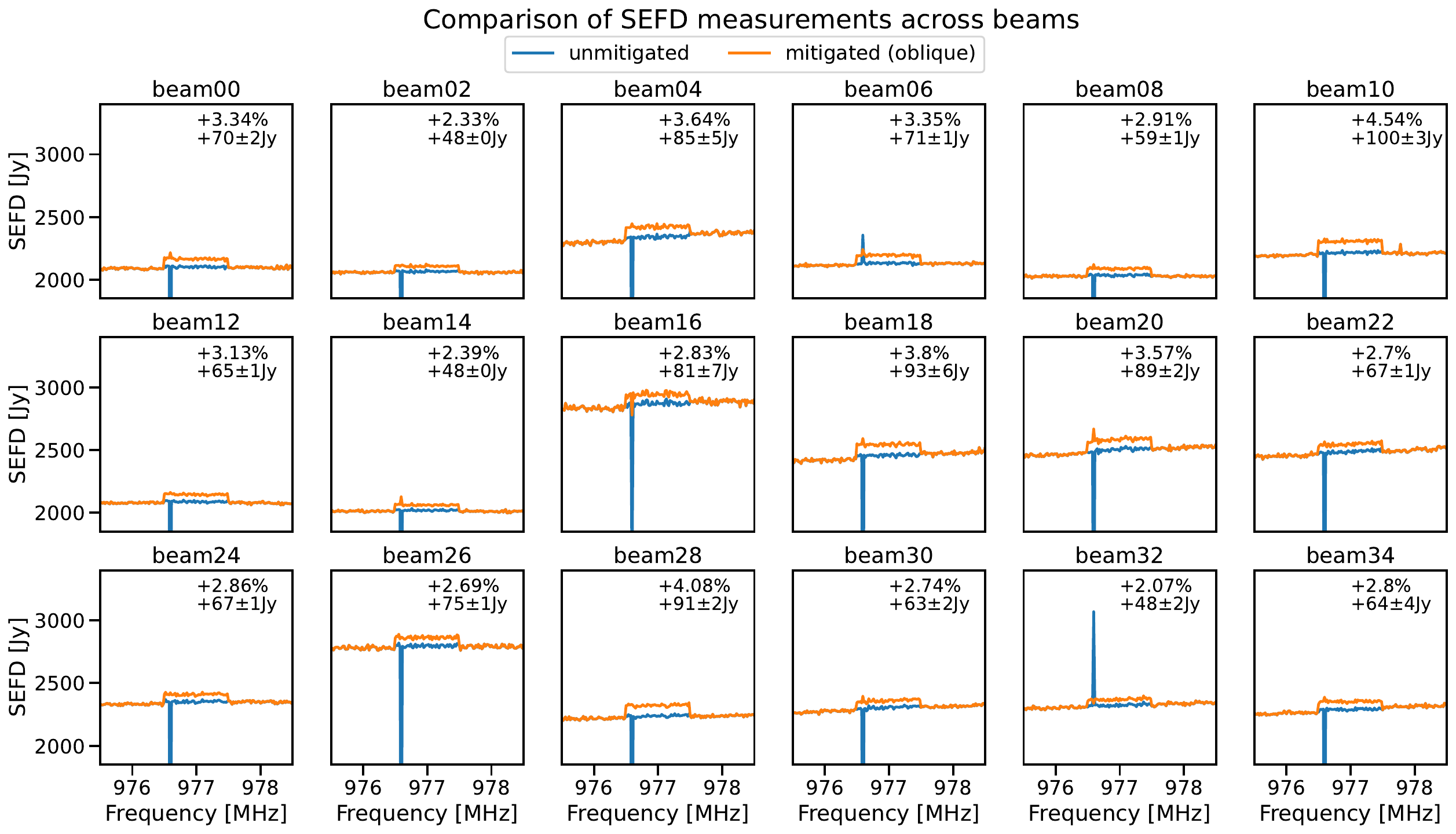}
    \caption{\ac{SEFD} measurement per beam, annotations showing the percentage increase and the increase in Janskys in the mitigated channel in orange introduced by the mitigation. The unmitigated \ac{SEFD} in blue. Each panel is averaged over all antennas.} 
    \label{fig:SEFD_allBeams}
\end{figure*}

Figure \ref{fig:SEFD_averaged} shows the \ac{SEFD} measurement with mitigation (orange) versus without mitigation (blue) averaged across all beams and antenna. The inset shows that similar to Figure \ref{fig:tsys}, the subspace projection introduces a measured \SEFDPercent~decrease in sensitivity in the mitigated channel. This increase in the system temperature is larger than what was seen in the fine filter bank test on Virgo in the previous experiment, and at the lower end of the \qtyrange{3}{5}{\percent} calculated value using the y-factor. Figure \ref{fig:SEFD_allBeams} separates out the individual beams (averaged over antennas) in the same \ac{SEFD} measurement. Each panel shows the corresponding pair of mitigated (orange) and unmitigated beams (blue) showing three adjacent \qty{1}{\mega\hertz} channels. In each channel containing the interferer, the offset in the mitigated channel is annotated. Offsets range from 48$\pm$0.8Jy to 100$\pm$3Jy, much less than the $\approx$800~Jy difference between beams. Compared to the calculated value using the correlation of the spatial signature vector and weights, the \ac{SEFD} values are higher than the mean calculated value of \qty{1}{\percent}, but all lower than the maximum calculated value of \qty{6}{\percent}.

\subsubsection*{Measured change in beamshape}
\begin{figure}[H]
    \centering
    \includegraphics[width=0.75\textwidth]{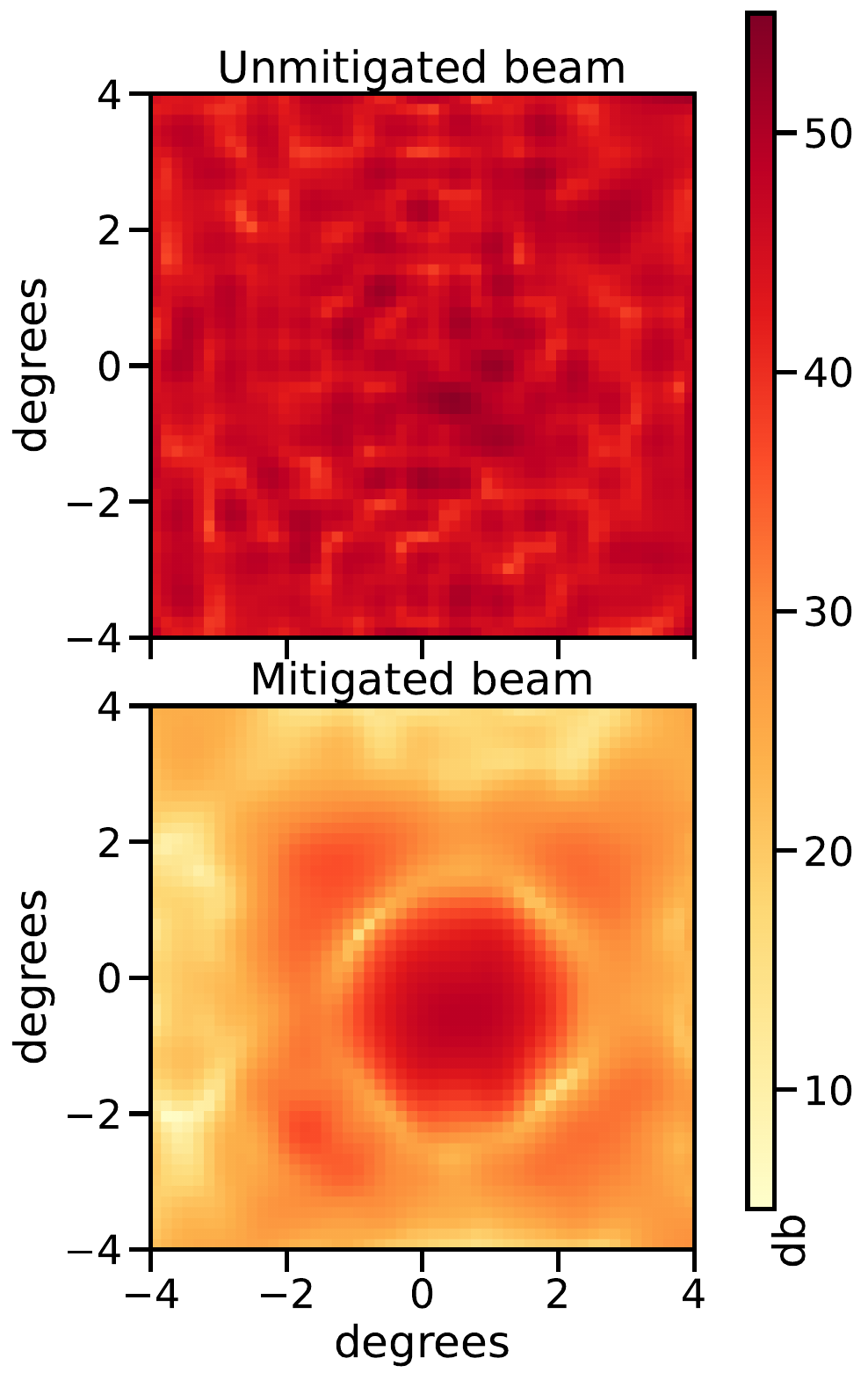}
    \caption{Holography measurement at \qty{977}{\mega\hertz} with mitigation (bottom) and without mitigation (top). Beams at this frequency are routinely corrupted by \ac{RFI} but can be completely recovered using the oblique projection.} 
    \label{fig:holo_singlebeam}
\end{figure}

Holography was used to determine changes to the beam shape and gain using the same footprint. Figure \ref{fig:holo_singlebeam} shows two beams located in position 0 of the footprint at \qty{977}{\mega\hertz} with (bottom) and without mitigation (top). Holography measurements are performed using a raster grid while observing a calibration source. In comparison to a reference boresight beam, the sensitivity pattern of all beams in the footprint is sampled \citep{Hotan2016HolographicBeams, Mcconnell2022ASKAPsMeasurement}. 

The top panel of Figure \ref{fig:holo_singlebeam} shows the effect of \ac{RFI} on holography.  The gross impact of \ac{RFI} on beamformer weights is already corrected using frequency interpolation \citep{Chippendale2017InterferenceTelescope}.  The degradation in the holography measurement is likely because the correlations are dominated by the correlating \ac{RFI} instead of the weaker astronomical source we are doing holography on. Not only is there an order of magnitude difference in the peak values between the two panels of Figure \ref{fig:holo_singlebeam}, but there is also no discernible beam in the top panel. Furthermore, this is not the case in the adjacent \qty{1}{\mega\hertz} channels free from \ac{RFI}. Comparison with archived holography measurements at \qty{977}{\mega\hertz} shows that beam maps at this frequency are regularly corrupted.

The mitigated copy of the beam is shown in the bottom panel of Figure \ref{fig:holo_singlebeam}. The mitigated beam pattern in Figure \ref{fig:holo_singlebeam} clearly shows the high-gain main lobe terminating in a surrounding null. We can also identify the side lobes surrounding the first null. 

To quantitatively measure differences in the mitigated beam pattern, we interpolate between the adjacent channels' sensitivity patterns to establish a reference beam to which to compare, assuming the channel-to-channel variations are smooth. We then fit a Gaussian to the reference `beam' to find the half power point, sample the reference beam and integrate around the half-power contour. We do the same for the mitigated beam and calculate the percentage difference between them at half power. 

\begin{landscape}
    \begin{figure}[ht]
        \centering
        \includegraphics[width=1.03\textwidth]{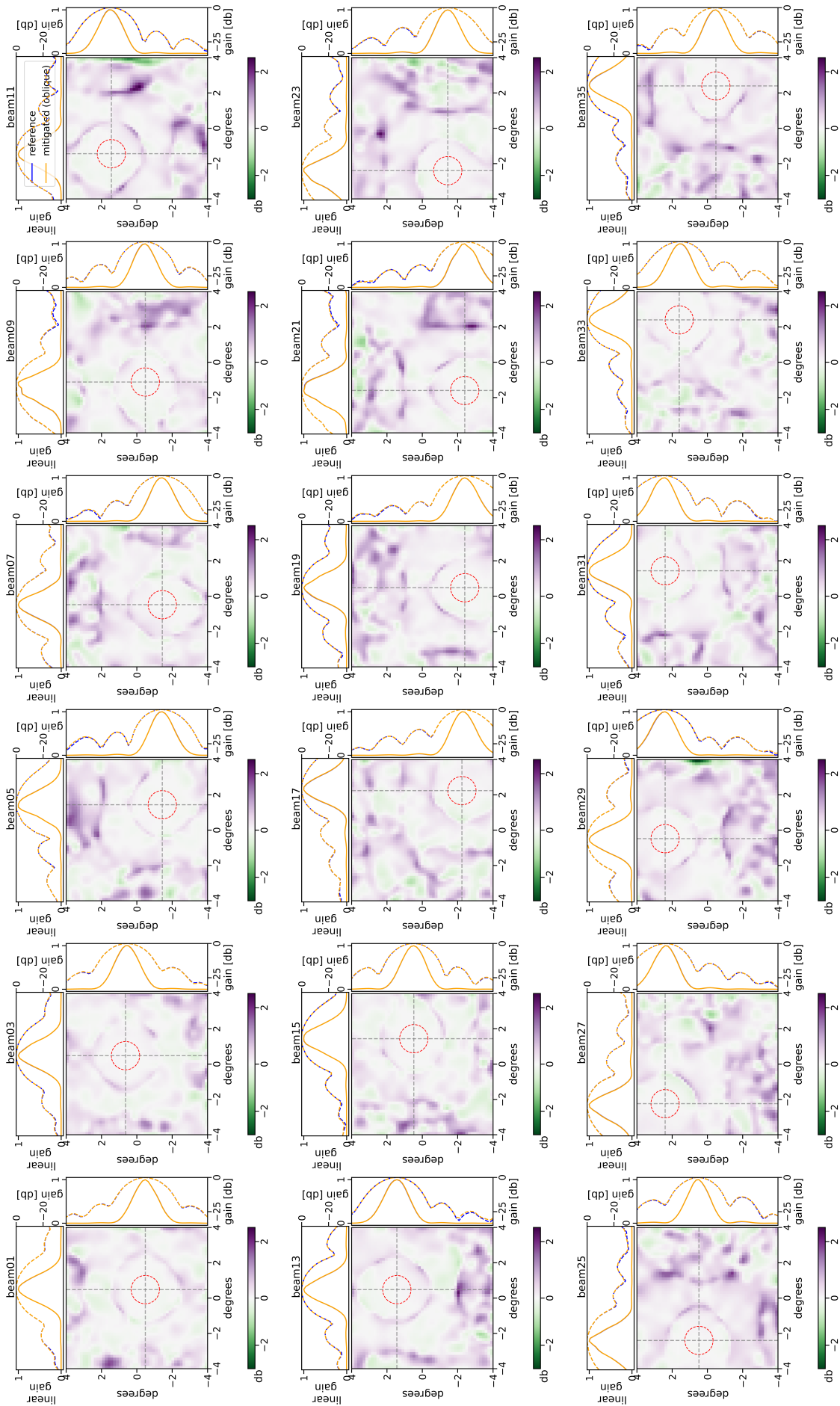}
        \caption{Comparison between the reference and mitigated beams by oblique projection across the modified close-pack square $6\times6$ footprint. Each beam has a central heatmap showing the difference between reference and mitigated beams in dB. The red dashed circle marks the half power point, white shows no change to the beam. Purple is an increase in gain, resulting from changes to the weights by introducing the green null (decrease in gain). All differences are within -\qtyrange{3}{2.5}{\decibel} and primarily limited to the outer lobes as predicted by the modelling. The top and side panels of each heatmap show a slice through the main beam at the grey dashed lines of each heatmap. These subplots show the reference beam (blue) and mitigated beam (overlaid in orange) with the normalized linear gain using solid curves and gain in dB using dashed curves.}
        \label{fig:holo_allbeams}
    \end{figure}
\end{landscape}

Figure \ref{fig:holo_allbeams} shows differences between the reference and mitigated beams across the footprint. The heatmaps in Figure \ref{fig:holo_allbeams} shows that there is no change (white) to the shape or gain in the main lobe. The heatmaps shows the difference between the reference and mitigated beams in dB. The half power point is marked by the red dashed circle. Green indicates suppression of the \ac{RFI} in the beam pattern by oblique projection. Purple, an increase in gain, is the compensation as a result of the mitigation of introducing the (green) suppression. The purple also represents the underlying reason for the increase in the system temperature. 

The subplots in Figure \ref{fig:holo_allbeams} (the top and right panels of each individual heatmap) show a slice across the main heatmap through the maximum of the beam identified by horizontal and vertical dashed grey lines, respectively similar to (and with the same colours as) Figure \ref{fig:beamshapeComparison}. The top and side panels show the reference beam in blue and the mitigated beam overlay in orange through that slice. The solid curves show the normalized linear gain. The dashed curves show the beamformed response in dB.

Two points worth noting; first, in Figure \ref{fig:holo_allbeams}, had the interferer been external to \ac{ASKAP}'s receivers, the suppressed direction of the interferer (Green) would have been `better defined'. Second, Figures \ref{fig:holo_singlebeam} and \ref{fig:holo_allbeams} have been interpolated to a higher resolution from the original holography measurements.

On average the mitigated beam differs from the main beam about the half power point by only \qty{0.02}{\percent}, with a range of just over \qty{1.5}{\percent} and standard deviation of \qty{0.4}{\percent}. Figure \ref{fig:holo_allbeams} shows that all differences are within \qtyrange{-3}{2.5}{\decibel} of the reference beam and, in line with our modelling, changes to the beam shape are limited to the outer sidelobes.

\section{Future Work}
We are in the process of making this operational on \ac{ASKAP}. We have also found a second self-generated \ac{RFI} signal in \ac{ASKAP} data at \qty{960}{MHz} that is persistently flagged in continuum observations \citep{Lourenco2024SurveyStatistics}. This is due to the ADC read-out clock at \qty{320}{\mega\hertz} being aliased by the \qty{1280}{\mega\hertz} sample clock when observing with the \qty{1200}{\mega\hertz} bandpass filter.  We can mitigate this second signal using the exact method presented above. 

We would like to provide users with information on which \ac{RFI} may be successfully mitigated without exceeding specific limits on performance degradation. During observations, these impact estimates may be provided to observers and observing systems \citep{Moss2022TheTelescopes} to inform whether activating the mitigation will benefit science goals. The calculations from the correlation of the beam weight steering vector and the \ac{RFI} spatial signature seem sufficient to predict changes to the \ac{RFI}-affected channel and can therefore be recommended to guide users.

Furthermore, calculations of the expected noise power from the correlation of an estimated RFI spatial signature using the full \ac{ACM} with maxSNR weights using all ports, as opposed to only 60, show potential for improved mitigation on \ac{ASKAP}.

This method, applying projected beam weights at the beginning of observation and leaving them fixed, would also work for a phased array pointing directly at the sky, which doesn't track the source where the interferer is nearby and ground-based. In this case, there is also no relative motion between the array elements and the interferer(s).  However, additional techniques to determine the \ac{RFI} subspace would be required, especially at low frequencies \citep{Hellbourg2012CyclostationaryAstronomy}. 

In order to mitigate moving interferers relative to the \ac{ASKAP} --- and in the future, the Murriyang (Parkes) cryogenic \citep{Dunning2023AAstronomy} --- \acp{PAF}, we are currently developing strategies to identify \ac{RFI} affected channels and determine their suitability for mitigation based on their eigenvalues (power), eigenvectors (spatial signature), and estimate performance impacts via orthogonality of the \ac{RFI} eigenvactors and the maxSNR beam weights (Lourenço \& Chippendale in prep).

\section{Conclusion}
In the growing presence of \ac{RFI}, sensitive synthesis array instruments will need to rely on methods other than flagging for \ac{RFI} mitigation. Mitigating self-generated interference by estimating the spatial signature of the interferer once at the beginning of the observation is a practical initial use case of \ac{ASKAP}'s \ac{RFI} mitigation capabilities via \ac{PAF} beamforming before implementing the dynamic case --- adjusting beam weights continuously throughout the observation --- required to mitigate \ac{RFI} from moving sources, e.g. signals from satellites and aircraft (or static sources with a tracking telescope). 

This paper presents the effects on gain, beampattern and sensitivity of \ac{RFI} mitigation using subspace-projection-based spatial nulling algorithms on a self-generated stationary clock signal. Our modelling and measurements show that the impact on the main beam shape and sensitivity is small. Measured results on sensitivity and beam impacts agree with theoretical calculations via $\rho$ (correlation between \ac{RFI} subspace and beam weights) as well as simulations using a geometric optics model of the \ac{RFI}. We have also shown a technique for estimating weak narrowband interference subspaces by taking the difference of the mean adjacent channels and the \ac{RFI}-impacted channel \acp{ACM}. Suppression of the unwanted signal by over \achievedSuppresion~, to our measurement noise floor, was achieved with a \tsysPercent~ measured average increase to the system-temperature-over-efficiency using a single antenna and a \SEFDPercent~ increase in the \ac{SEFD} across the array. We modelled and measured no significant change to the gain in the main beam, \holoDifference~ on average, about the half power point. Variations to the beam pattern were limited to the outer lobes.

\section*{Acknowledgments}
We would like to thank Dr. Gregory Hellbourg and Professor Tara Murphy for their helpful feedback. We would also like to thank Dr. Tim Galvin for helping us run the \ac{SEFD} processing on supercomputers. 

This scientific work uses data obtained from Inyarrimanha Ilgari Bundara, the \acs{CSIRO} \acl{MRO}. We acknowledge the Wajarri Yamaji People as the Traditional Owners and native title holders of the observatory site. CSIRO’s ASKAP radio telescope is part of the Australia Telescope National Facility (https://ror.org/05qajvd42). Operation of ASKAP is funded by the Australian Government with support from the National Collaborative Research Infrastructure Strategy. ASKAP uses the resources of the Pawsey Supercomputing Research Centre. Establishment of \ac{ASKAP}, Inyarrimanha Ilgari Bundara, the \acs{CSIRO} \acl{MRO} and the Pawsey Supercomputing Research Centre are initiatives of the Australian Government, with support from the Government of Western Australia and the Science and Industry Endowment Fund.

\bibliography{references}

\appendix
\section{Effects of changes in phase on spatial nulling}
\label{appendix::coherent_vs_incoherent}

\begin{figure}[h]
    \centering
    \includegraphics[width=0.95\textwidth]{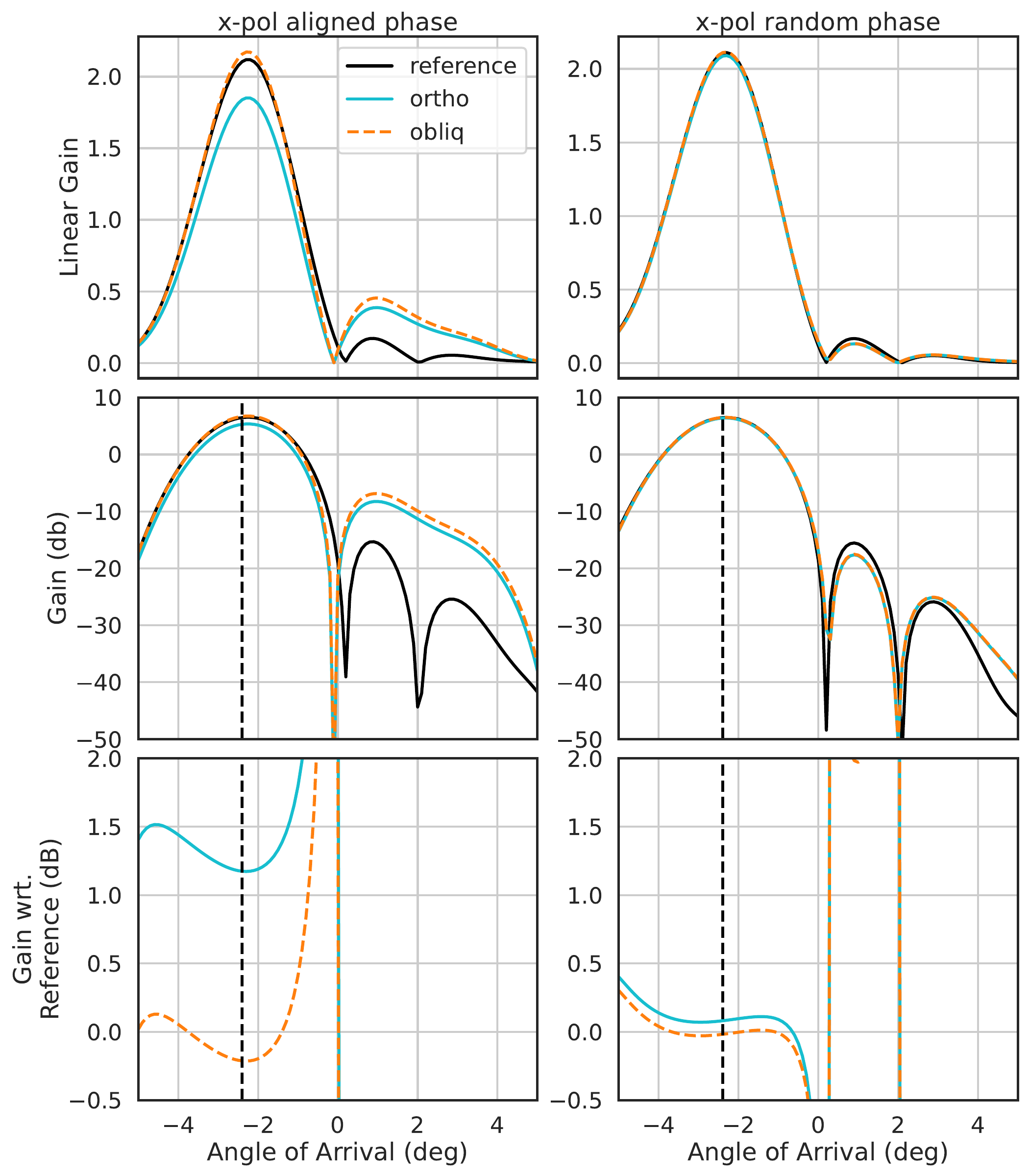}
    \caption{X polarisation beamformed response in the direction \qty{-2.4}{\degree}, before and after mitigation, showing that variations to the beampatterns are reduced when the phase of the generated signal is randomised. The rows show the linear gain, gain in decibels and the percentage difference in linear gain compared to the unmitigated beam.}
    \label{fig:CoherentvsIncoherent}
\end{figure}

In the process of constructing the model, we considered two instances of self-generated \ac{RFI}; a uniform amplitude arriving at all ports with and without a random phase, i.e. randomised and aligned phase, respectively. Figure \ref{fig:CoherentvsIncoherent} shows an X polarisation beam centred in the direction \qty{-2.4}{\degree}, before and after mitigation. 

Changes to the beam based on an \ac{RFI} spatial signature with aligned phase on the left and randomised phase on the right are shown using both the orthogonal and oblique projections.  The rows show the linear gain, gain in decibels and the percentage difference in linear gain to the unmitigated beam respectively. After mitigation, more significant variations to the beam patterns were observed in both the main lobe and the outer lobes without a uniform phase compared to a random phase, where variations were most pronounced starting at the second null and an almost complete agreement between the reference and projected beamformed patterns in the main lobe and first side lobe. Although, the top left panel of Figure \ref{fig:CoherentvsIncoherent} shows that the oblique projection corrects for the error in the main beam experienced by the orthogonal projection when the \ac{RFI} subspace is randomised in phase. Variations to the beampatterns are reduced when the phase of the generated signal is randomised. The measured \ac{RFI} spatial signature in Figure \ref{fig:beamshapeComparison} is consistent with a clock signal arriving at the PAF ports with differing delays along the signal path of each port.  The delay differences are dominated by random digital delays that are introduced each time \ac{ASKAP}'s digital receivers are reset.

\end{document}